# Momentum-dependent sign-inversion of orbital polarization in superconducting FeSe

Y. Suzuki[1], T. Shimojima[1], T. Sonobe[1], A. Nakamura[1], M. Sakano[1], H. Tsuji[1],
J. Omachi[2], K. Yoshioka[3], M. Kuwata-Gonokami[2,3], T. Watashige[4], R. Kobayashi[4], S. Kasahara[4],
T. Shibauchi[5], Y. Matsuda[4], Y. Yamakawa[6], H. Kontani[6], K. Ishizaka[1]

[1]*Quantum-Phase Electronics Center (QPEC) and Department of Applied Physics, University of Tokyo, Bunkyo, Tokyo 113-8656, Japan*
[2]*Photon Science Center, The University of Tokyo, 7-3-1 Hongo, Bunkyo-ku, Tokyo 113-8656, Japan*
[3]*Department of Physics, The University of Tokyo, 7-3-1 Hongo, Bunkyo-ku, Tokyo 113-0033, Japan*
[4]*Department of Physics, Kyoto University, Kyoto 606-8502, Japan*
[5]*Department of Advanced Materials Science, University of Tokyo, Chiba 277-8561, Japan*
[6]*Department of Physics, Nagoya University, Furo-cho, Nagoya 464-8602, Japan*

We investigate the electronic reconstruction across the tetragonal-orthorhombic structural transition in FeSe by employing polarization-dependent angle-resolved photoemission spectroscopy (ARPES) on detwinned single crystals. Across the structural transition, the electronic structures around the $\Gamma$ and $M$ points are modified from four-fold to two-fold symmetry due to the lifting of degeneracy in $d_{xz}/d_{yz}$ orbitals. The $d_{xz}$ band shifts upward at the $\Gamma$ point while it moves downward at the $M$ point, suggesting that the electronic structure of orthorhombic FeSe is characterized by a momentum-dependent sign-changing orbital polarization. The elongated directions of the elliptical Fermi surfaces (FSs) at the $\Gamma$ and $M$ points are rotated by 90 degrees with respect to each other, which may be related to the absence of the antiferromagnetic order in FeSe.

Most of the parent compounds of the iron-based superconductors show the tetragonal-orthorhombic structural transition at $T_s$ and the stripe-type antiferromagnetic (AFM) order below $T_N$ ($\leq T_s$) [1,2]. Near the structural transition, an orbital order defined by the inequivalent electron occupation of $3d_{xz}$ ($xz$) and $3d_{yz}$ ($yz$) orbitals [3-5], has been reported by ARPES [6,7] and X-ray linear dichroism measurements [8] in several parent compounds. Experimental and theoretical studies suggested that the structural transition is caused by the electronic nematicity of the spin [9,10] or orbital [11-13] degrees of freedoms. Since superconductivity develops when such complex ordered states are suppressed, it is crucial to understand how the phase transitions couple to each other.

In Ba(Fe,Co)$_2$As$_2$, the spin-driven nematicity has been suggested from the phase diagram in which $T_s$ and $T_N$ closely follow each other as the carrier is doped [14]. The scaling behavior between the nematic fluctuation and spin fluctuation was also reported by the nuclear magnetic resonance (NMR) and shear modulus measurements [10]. On the other hand, in NaFeAs, the orbital-driven nematicity has been proposed by ARPES [11]. In this compound, the structural transition at $T_s = 54$ K is well separated from the AFM transition at $T_N = 43$ K. Inequivalent shift in the $xz/yz$ orbital bands appearing above $T_s$ changes the FSs from four-fold to two-fold symmetric shape [11,15], which may be a possible trigger of the stripe type AFM order and the orthorhombicity [11,16]. The variety of iron-based superconductors thus requires us to investigate how the driving force of the electronic nematicity depends on each material class.

FeSe is a good example to examine the role of orbital degrees of freedom, since it shows the structural and superconducting (SC) transitions at $T_s \sim 90$ K and $T_c = 9$ K without any magnetic order [17]. Recent ARPES studies on FeSe reported a lifting of degeneracy in $xz/yz$ orbitals at the $M$ point showing up at low temperature [12,18-21]. Since the difference in the energy between $xz$ and $yz$ bands is much larger than that expected from the orthorhombic lattice distortion alone, it is considered to be an indication of the electronic nematicity of orbital-origin [12]. It is further expected that the unconventional superconductivity with nodal superconducting gaps is realized in FeSe [22,23] on the two-fold FSs in the orbital ordered state. While a drastic modification of the FSs has been reported by ARPES on twinned crystals [19], there has been no report directly showing the two-fold symmetry of the FS shape and its orbital character in the single domain. In order to understand the orbital-driven nematicity and its relation to the superconductivity, we must clarify how the orbital order affects the electronic structure in the entire Brillouin zone (BZ) by using the detwinned crystals of FeSe.

In this Letter, we report the electronic reconstruction across $T_s$ of FeSe by using He-discharge lamp and laser-ARPES on detwinned crystals.

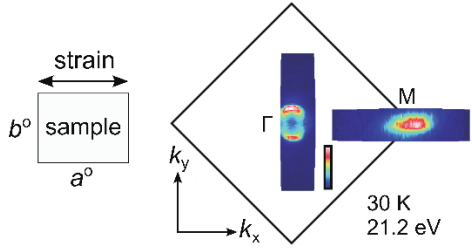

**Figure 1**. Fermi surface mapping taken at 30K by $h\nu$ = 21.2 eV. Black square represents the tetragonal Brillouin zone. $x$ and $y$ are coordinates along the crystal axes of the orthorhombic setting $a^o$ and $b^o$, respectively ($a^o > b^o$). The $a^o$ axis is aligned along the direction of the uniaxial tensile strain.

We observed a drastic modification of FSs from four-fold to two-fold symmetry around both the $\Gamma$ and $M$ points. The electron FS at the $M$ point and the hole FS at the $\Gamma$ point were found to be of elliptical shape rotated by 90 degrees with respect to each other. Polarization-dependent ARPES further revealed that the $xz$ orbital shifts upward at the $\Gamma$ point while it moves downward at the $M$ point, thus indicating a momentum-dependent sign-inversion of orbital polarization in $xz/yz$ orbitals. Imperfect FS nesting condition in the orthorhombic phase might be related to the absence of the antiferromagnetic order in FeSe.

High quality single crystals of FeSe were synthesized as described in ref.17. The transition temperatures of the single crystals were estimated to be $T_s \sim 90$ K and $T_c \sim 9$ K from the electrical resistivity measurements. For the FS mapping below $T_s$, we used a VG-Scienta R4000WAL electron analyzer and a helium discharge lamp of $h\nu = 21.2$ eV at the University of Tokyo. The energy resolution was set to 15 meV. For the polarization-dependent ARPES at the $\Gamma$ point, we employed the fourth-harmonic generation of Ti:sapphire laser radiation ($h\nu = 5.9$ eV) [24] and a VG-Scienta R4000WAL electron analyzer at the University of Tokyo. We chose $s$ and $p$ polarizations to determine the orbital characters. The energy resolution was set to be 6 meV. To detwin the single crystals, we applied an in-plane uniaxial tensile strain, which brings the orthorhombic $a^o$ axis ($a^o > b^o$) along its direction below $T_s$ (see Fig. 1). Samples were cleaved *in situ* at room temperature in an ultrahigh vacuum of $5 \times 10^{-11}$ Torr.

Shown in Fig. 1 is the FS mapping with an energy window of ±5 meV taken at 30 K ($T < T_s$) for detwinned FeSe. In the orthorhombic phase, we found one elliptical hole FS at $\Gamma$ and one elliptical electron FS at $M$, with the long axes along $k_y$ and $k_x$, respectively. According to our previous ARPES on detwinned FeSe [12], elliptical intensity at the $M$ point is mainly derived from the electron bands composed of $xz$ orbital along $k_x$ and $yz$ orbital along $k_y$ direction. It is in a strong contrast to the high temperature tetragonal phase, where a small circular hole FS at $\Gamma$ and two crossing elliptical electron FSs at $M$ are observed in the four-fold symmetry [12].

In order to assign the orbital characters of the band dispersions which make the hole FS around the $\Gamma$ point, we employed laser-ARPES with $s$- and $p$-polarizations. Considering the parity symmetry of light polarizations with respect to a mirror plane, one can assign the orbital character for each band dispersion [25]. In Figs. 2(a) and 2(b), we illustrate the experimental geometries for different sample orientations. In Geometry 1 where the strain direction is parallel to the detector slit, the momentum along the short axis of the elliptical FS is measured. Here, $yz$ and $xy$ ($xz$) orbitals have odd (even) symmetry with respect to the mirror plane and can be detected by $s$ ($p$)-polarized light. In a similar way, $xz$ and $xy$ ($yz$) are supposed to be detected by $s$ ($p$)-polarized light in Geometry 2 which measures along the long axis of the elliptical FS.

Figs. 2 (c-e) show an $E$-$k$ image, its second $E$ derivative image and the schematic band dispersion obtained by laser-ARPES in Geometry 1 (along $k_x$) at 160 K, respectively. The left and right panels are the results taken by $s$- and $p$-polarized light, respectively. In Figs. 2(c) and (d), one can find that the $\alpha$ band crosses $E_F$ at the Fermi momentum ($k_F$) of ~ 0.07 Å$^{-1}$ while the top of the $\beta$ band locates at $E - E_F \sim -10$ meV. In Geometry 2, $\alpha'$ and $\beta'$ hole bands are similarly observed as shown in Figs. 2(f) and 2(g). Considering the polarization dependence in Geometries 1 and 2, $\alpha$ and $\beta'$ ($\beta$ and $\alpha'$) bands are consistently assigned to be dominantly of $yz$ ($xz$) orbital character. Relatively flat band of low intensity around $E - E_F \sim -50$ meV only detected by $s$-polarization originates from $xy$ orbital component. Schematic band structures in the tetragonal phase are then summarized in Fig. 2 (e and h). These results suggest that the circular hole FS around the $\Gamma$ point is equally composed of $xz$ and $yz$ orbital characters.

ARPES images taken at 30 K show a drastic difference from the band dispersions at the tetragonal phase. In Figs. 2(i) and 2(j), we clearly observed three hole-like bands ($\alpha$, $\beta$, $\gamma$) along $k_x$. The $\alpha$ band crosses $E_F$ at $k_F \sim 0.04$ Å$^{-1}$, while the $\beta$ and $\gamma$ bands show their tops at $E - E_F \sim -20$ meV and ~ -50 meV, respectively. As summarized in Fig. 2(k), the $\alpha$ band has $xz$ orbital character near the $\Gamma$ point, which seems to gradually change into $yz$ near $k_F$. One can thus notice that the orbital component of the $\alpha$ band around $E_F$ is partly changed from $yz$ to $xz$ across $T_s$. In Geometry 2, as shown in Figs 2(l) and 2(m), we observed the $\alpha'$ and $\gamma'$ hole bands in the left panel ($s$-polarization) and $\beta'$ and $\gamma'$ bands in the right panel ($p$-polarization). The $\alpha'$ band crosses $E_F$ with $k_F \sim 0.10$ Å$^{-1}$, while the $\beta'$ and $\gamma'$ hole bands show their tops at $E - E_F \sim -20$ meV and ~ -50 meV, respectively. In Fig. 2(n), we summarized the band dispersions and orbital characters along $k_y$. Figures 2(k) and 2(n) show that the inequivalent band shift in $xz/yz$ orbitals occurs around the $\Gamma$ point, resulting in $E_{xz} > E_{yz}$. Considering that $E_{yz} > E_{xz}$ is realized at the $M$ point [12], we can identify the opposite sign of the orbital polarization realized for the $\Gamma$ and $M$ points. It is worth mentioning that the hole FS is mainly composed of $xz$ orbital along $k_y$, and $xz$ and $yz$ orbitals along $k_x$. The $k_F$ along $k_x$ ($k_y$) is estimated to be 0.04 Å$^{-1}$ (0.10 Å$^{-1}$) being consistent with the elliptical FS shown in Fig.1.

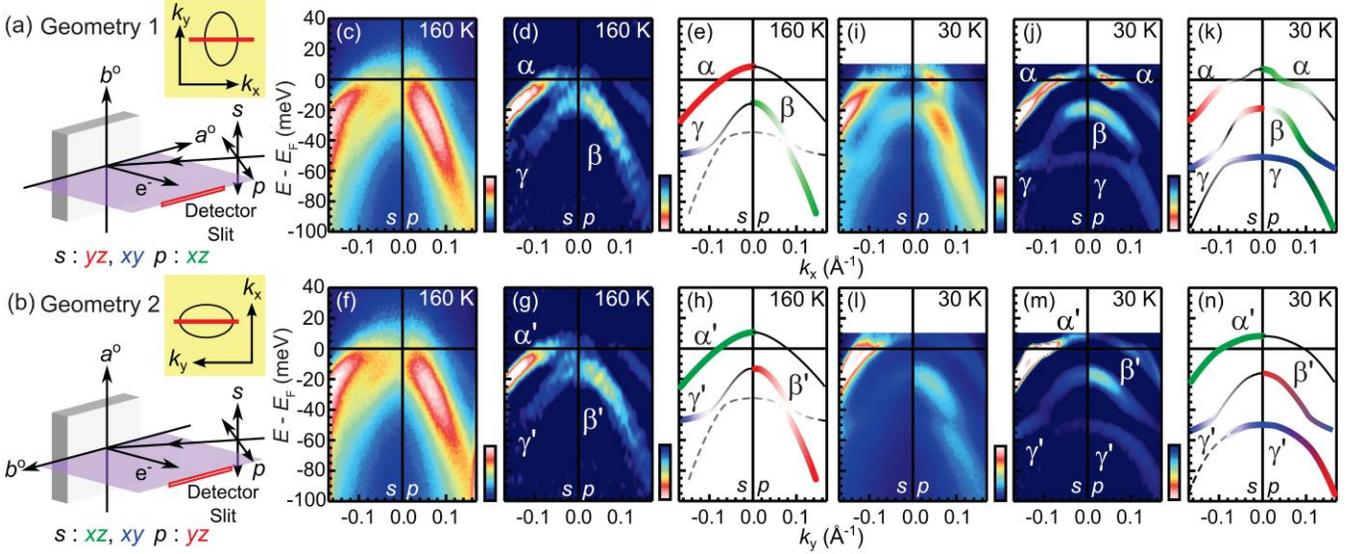

**Figure 2.** (a,b) Two experimental geometries for polarization-dependent laser-ARPES. In Geometry 1 (2), orthorhombic $a^o$ axis is parallel (perpendicular) to the detector slit, owing to the detwinning procedure. Purple plane represents a mirror plane of the orthorhombic lattice. Observable orbital characters are also shown for each polarization. (c,d) $E$-$k$ image divided by FD function and its second $E$ derivative detected in Geometry 1 at 160 K. The left (right) panels are $E$-$k$ images obtained by $s(p)$-polarization. (e) Schematic band dispersions and their orbital characters. Band dispersions colored in green, red and blue are composed of $xz$, $yz$ and $xy$ orbitals, respectively. The black curves are the guides for the eyes. The broken curves represent the guides showing the band dispersions which were not clearly observed by ARPES. (f-h), The same as panel (c)-(e) taken in Geometry 2 at 160 K. (i-k), The same as panel (c)-(e) taken in Geometry 1 at 30 K. (l-n) The same as panel (c)-(e) taken in Geometry 2 at 30 K.

In order to interpret the electronic reconstruction around the $\Gamma$ point, we compared the present ARPES results with the band calculations. We constructed the tight-binding model based on the DFT calculations using WANNIER90 code and WIEN2WANNIER interface. Lattice parameters used for the band calculations were taken from Ref. [26]. Then, we included the spin-orbit interaction (SOI) with $\lambda$ = 50 meV, by which the degeneracy of the $xz/yz$ bands at the $\Gamma$ point is lifted. Next, according to the previous ARPES report [27], we renormalized the band structure with the renormalization factors of ~ 3 for the $xz/yz$ orbitals and ~ 9 for the $xy$ orbital. Finally, realistic tight-binding model above $T_s$ is obtained by shifting the $xz/yz$ and $xy$ bands as in Ref. [28]. Furthermore, in order to reproduce the orbital ordered state below $T_s$, we introduced the orbital polarization at the $\Gamma$ point by shifting the $xz$ ($yz$) band by +7 meV (-7 meV). The sign of these band shifts are opposite to those at the $M$ point.

In the tetragonal phase, the present laser-ARPES results summarized in Fig. 3(a) are well reproduced by the band structure of the realistic tight-binding model in Fig. 3(c). The band structure is four-fold symmetric, but $\alpha$, $\beta$ bands are not degenerate at the $\Gamma$ point, showing an energy gap of 20 meV. This gap scale is consistent with the previous ARPES [19,20] and here we can explain it by the effect of SOI. Kink structures in the bands $\beta$ and $\beta'$ at $E - E_F$ ~ -50 meV are also possibly an indication of the small gapped feature due to SOI as shown in Fig. 3(c).

In the orthorhombic phase, the calculations in Fig. 3(d) capture some important features in the present ARPES results in Fig. 3(b). Large anisotropy of the $k_F$ is reproduced by the effect of the orbital order of $E_{xz} > E_{yz}$. The energy gap between $\beta$ and $\gamma$ bands is clearly observed owing to the high-energy resolution of the laser-ARPES, and considered to be a signature of the SOI. We note that the splitting energy between $\alpha$ and $\beta$ bands at the $\Gamma$ point becomes ~ 10 meV larger than that in the tetragonal phase. We thus conclude that the complex band dispersions in Fig. 3(b) are explained by the $T$-independent component from SOI (~ 20 meV) and $T$-dependent one due to the orbital order (~ 10 meV). The present result shows that in the orthorhombic phase in FeSe, the orbital polarization $f(k) = E_{yz}(k) - E_{xz}(k)$ strongly depends on $k$, and it becomes negative around the $\Gamma$ point (-10 meV). Since the magnitude of the orbital polarization is much larger at the $M$ point (+50 meV), the relation $n_{xz} > n_{yz}$ should be satisfied similarly to other compounds.

Figures 4 (a) and (b) summarize the FSs of FeSe obtained by the detwinned ARPES above and below $T_s$. Schematic FSs around the $\Gamma$ and M point are obtained from the ARPES data taken by hν = 5.9 eV and 60 eV [12], respectively. In contrast to the four-fold symmetric FS in the tetragonal phase, both the shape and the orbital component of the FSs become two-fold symmetric below $T_s$. We can see one elliptical FS sheet around the $\Gamma$ with dominantly $xz$ orbital component. Previous ARPES on twinned FeSe reported that the two elliptical hole FSs are crossing at the $\Gamma$ point. Owing to the detwinning procedure, here we clearly separated them and determined the orientation of the elliptical hole FS which is elongated along $k_y$.

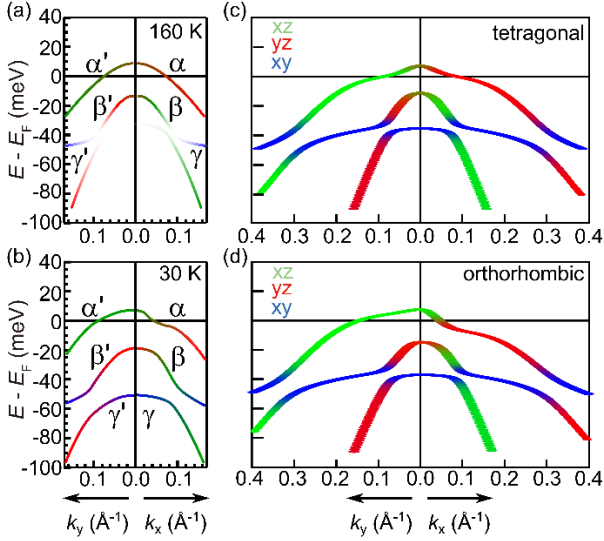

**Figure 3.** (a,b) Schematic band dispersions obtained by laser-ARPES at 160 K and 30 K, respectively. The left (right) panels are the results along $k_y$ ($k_x$). (c,d) The band dispersions of the realistic tight-binding model including the effects of SOI and orbital order. Band dispersions colored in green, red and blue are composed $xz$, $yz$ and $xy$ orbital components, respectively.

Regarding the FS at the $M$ point, we have reported highly anisotropic $k_F$ in the $xz$ and $yz$ electron bands arising from the orbital order of $E_{yz} > E_{xz}$ [12]. Elliptical intensity around the $M$ point in the FS mapping in Fig. 1 is consistent with this picture of anisotropic $k_F$. The FS corresponding to the $xy$ band as detected in the tetragonal phase [12], on the other hand, is not clearly observed in the orthorhombic phase [depicted by the broken curves in Fig. 4(b)].

Our observation of the two elliptical FSs with different orientations at the $\Gamma$ and $M$ point is consistent with the quasiparticle interference (QPI) observed by STM/STS on FeSe [23]. As reported in ref. 23, hole (electron) band-like dispersion in the QPI image is clearly observed only along $q_a$ ($q_b$) ($a > b$). Such anisotropy in the QPI images is understood by considering that the QPI is dominant between the flat portions of respective elliptical FSs at the $\Gamma$ and $M$ points.

It is also worth mentioning that the lifting of degeneracy of $E_{yz} > E_{xz}$ at the $M$ point with its orbital polarization larger than that at the $\Gamma$ point is common to several iron-based superconductors. These facts indicate that the relation in the occupied electron numbers of $n_{xz} > n_{yz}$ always holds, which might be an origin of the orthorhombicity. Theoretical studies suggested that the $k$-independent orbital polarization on the order of 10 meV resolves the magnetic frustration between $\mathbf{q} = (\pi,0)$ and $(0,\pi)$ [16]. For FeSe, actually, strong spin fluctuations were observed for $\mathbf{q} = (\pi,0)$ by the neutron scattering measurement [29] and its enhancement below $T_s$ was reported by NMR measurement [30], possibly reflecting the above scenario.

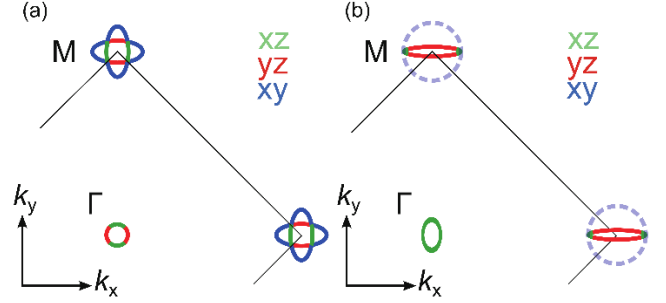

**Figure 4.** (a,b) Schematic FSs obtained by ARPES on detwinned FeSe in the tetragonal phase and orthorhombic phase, respectively. FSs colored by green, red and blue are composed of $xz$, $yz$ and $xy$ orbitals, respectively. Broken curve in (b) represents the electron FS sheets composed of $xy$ orbital expected to show up, but not observed in the present ARPES. Black square represents the tetragonal Brillouin zone.

However, it is puzzling that FeSe does not exhibit any AFM order despite the strongly enhanced spin fluctuation. Here we focus on the orbital polarization of $E_{xz} > E_{yz}$ around the $\Gamma$ point, which has been detected only for FeSe until now. As clearly shown in this work, the $k$-dependent sign-inversion of orbital polarization leads to the situation that the elliptical FS at the $\Gamma$ point (long-axis along $k_y$) is rotated by 90 degrees with respect to that at the $M$ point (long-axis along $k_x$). Such electronic reconstruction makes the FS nesting condition worse. In the case of the orthorhombic/paramagnetic phase ($T_N < T < T_s$) of NaFeAs, both of the elliptical FSs at the $\Gamma$ and $M$ points are aligned with their long-axes along $k_x$, thus making the FS nesting condition better [11,15], which is interpreted as the driving force of the AFM order [11]. Such difference from NaFeAs may be related to the absence of the AFM order in FeSe. We can thus expect that the AFM order in FeSe is suppressed due to the imperfect FS nesting between the $\Gamma$ and $M$ points, while $(\pi,0)$ spin fluctuations are substantially enhanced due to the magnetic frustrations resolved by the orbital ordering at $T_s$. Further investigations on the superconducting gap in the highly two-fold symmetric electronic structure in FeSe will give us fundamental information regarding the relation between the orbital ordering and the superconductivity in iron-based superconductors.

In conclusion, polarization-dependent ARPES study on detwinned FeSe revealed the drastic reconstruction of the FSs across $T_s$. The electronic structure in the orthorhombic phase is qualitatively explained by the *T*-independent component from the SOI and *T*-dependent part due to the orbital order. We found that the *xz* band lifts up around the *Γ* point while the *yz* band moves up around the *M* point. Such momentum-dependent sign-inversion of orbital polarization modifies the FSs at the *Γ* and *M* point into elliptical shapes elongating along $k_y$ and $k_x$, respectively. This modification makes the imperfect FS nesting condition, which may be related to the absence of the AFM state and electronic nematicity of orbital-origin in FeSe.